# Optical models for thin layers

Yilei Li and Tony F. Heinz *

*Department of Applied Physics, Stanford University, Stanford, California 94305, USA*
*SLAC National Accelerator Laboratory, Menlo Park, California 94025, USA*

* To whom correspondence should be addressed: tony.heinz@stanford.edu

**ABSTRACT**

We provide a systematic study of the optical models for thin layers: the 3D model, the 2D model and the linearized 2D model. We show that the 2D model is applicable for layers with small optical thicknesses. Excellent agreement of the 2D model with the 3D model is demonstrated over broad spectral ranges from DC to near UV for representative van der Waals atomic layers and thin metal layers. The linearized 2D model requires additionally weak optical response. Analytical expressions for the applicability and accuracies of the optical models are derived. We discuss the advantages and limitations of the models for the purpose of measuring their optical response functions. Further, we generalize the theory to take into account in-plane anisotropy, heterostructures, and stratified substrates. Implications of the 2D model for the correct analysis of transmission attenuation and implementations of half- and total-absorption layers are discussed.

# I. INTRODUCTION

During the past decade, the advent of atomically thin materials has generated great interest in the measurement and understanding of their optical properties. From the reflection contrast of graphene on a fused silica substrate, the imaginary part of the dielectric function was measured. [1] In a transmission configuration, the imaginary part of the dielectric function was obtained from the reduction in optical transmission through a suspended graphene layer. [2] Both of these methods applied a model in which the measured light intensity varies linearly with the imaginary part of the dielectric function. The full dielectric function is determined by an optimization procedure constrained by the Kramers-Kronig relation between the real and imaginary parts of the susceptibility.[3-5] In addition, ellipsometry has been applied to determine the optical response functions of atomically thin layers from the polarization state of the reflected light. [6-14]

In the earlier studies, the atomically thin materials are treated either as thin 3D slabs that are described by volumetric responses [3,4,6-14] or as 2D sheets described by sheet responses [1,2]. In certain cases, they are also treated as perturbations. [1,2] These models have been applied beyond the 2D materials. The 3D model has been the basic tool for designing thin film coatings [15]. The 2D model has been widely applied to model thin layers, too. [16,17] Applications include the development of metal film attenuators in the microwave region [18], the development of absorptive anti-reflection coating for radio waves [19], in the study of superconducting films in the terahertz and far infrared region [20-22], and in nonlinear optical studies of surfaces and interfaces. [23,24]

We examine the optical models systematically in this paper. In Section II, we layout the basic equations for the 2D and 3D models, as well as the linearized version of the 2D model. In Section III, we discuss approaches to extract the response functions from measurements. Scenarios where results can be obtained by directly inverting data will be described. In Section IV, we derive the analytical criteria for the applicability of the models and we present a set of criteria to infer the applicability directly from experimental observables. In Section V, we discuss the accuracy of the 2D model using representative 2D materials. We show that the 2D model is generally applicable and accurate for atomic layers in wide frequency ranges from DC to near UV and the linear model is appropriate with small optical contrast as a practical criterion. A thin gold layer is used as an example to show that the 2D model can be accurate even when the optical transmission is diminishing. In Section VI, we resolve an apparent contradiction with the Beer's law when analyzing the transmission attenuation of a thin layer and conditions for half- and total-absorbance are obtained. In Section VII, we generalize the models for materials with anisotropic in-plane response and for heterostructures.

## II. FORMULATION OF THE OPTICS OF A THIN LAYER

### II A. DESCRIPTION OF THE OPTICS PROBLEM

The general problem that we wish to address involves a thin layer of material lying on a semi-infinite substrate, as illustrated in Fig. 1. The materials are treated within a continuum approximation as smooth, homogeneous media, exhibiting optical responses that can be described by local dielectric functions. We assume for simplicity that in-plane response of the different media is isotropic. In the present analysis, we restrict our attention to the behavior for light approaching the system at normal incidence. The effect of the substrate and overlayer on the incident optical radiation is described by the complex reflection $r$ and transmission $t$ coefficients, given, respectively, as the ratios of the reflected and transmitted fields to the incident field. The goal of the theory is to obtain from the experimentally measureable quantities, such as $r$ and $t$ or their intensity analogs, the dielectric response of the layer and *vice versa*. Within the context of our discussion, we assume that optical response of the substrate, described by complex index $n_s$, is known. In the main text, we limit our treatment to the case of vacuum as a superstrate above the layer. The extension to the case of an arbitrary transparent superstrate is provided in the Supplemental Material, Section VI.

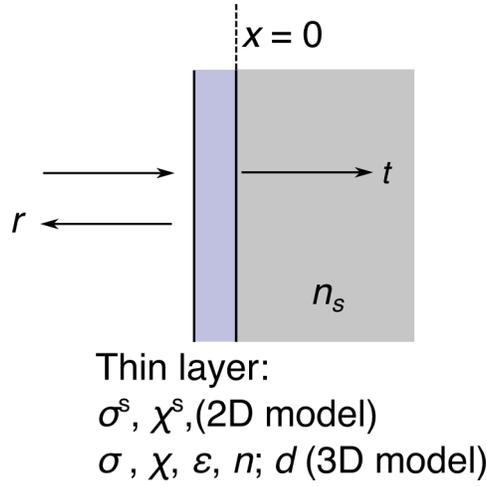

**Figure 1.** A thin layer of material lies between vacuum and a substrate with (complex) refractive index $n_s$. The optical response of the thin layer of thickness $d$ is described by the volumetric response functions ($\sigma$, $\chi$, $\varepsilon$, or $n$) or by its sheet response functions ($\sigma^s$ and $\chi^s$) in the 2D model. We consider the case of normally incident light and define the reflection $r$ and transmission $t$ coefficients using a coordinate system whose origin is at the surface of the substrate ($x = 0$).

Within the conventional 3D model of the system, the thickness $d$ of the overlayer is a key parameter. The optical response of this layer is characterized by any of the four equivalent volumetric response functions: the conductivity $\sigma$, the susceptibility $\chi$, the dielectric function $\varepsilon$ (the electric permittivity normalized by the vacuum permittivity), and the complex refractive index $n$. In the 2D model developed below, the overlayer is treated as a homogeneous sheet of negligible thickness. The optical response is then described, by the equivalent sheet response functions $\sigma^s$ and $\chi^s$, which are given by the product of the corresponding volumetric response functions and the layer thickness $d$. Table 1 summarizes the inter-related response functions of the overlayer in the 3D and 2D pictures.

**Table 1.** List of optical response functions and their relations. $\varepsilon_0 \approx 8.85 \times 10^{-12}$ F/m is the vacuum permittivity, $\omega$ is the angular frequency of light and $d$ is the layer thickness. Throughout this paper we adopt the SI units and the sign convention is discussed in the Supplemental Information, Section I.

|  | Optical response functions |  | Relations |
|---|---|---|---|
| Volumetric response functions | Conductivity | $\sigma$ | $-i\chi\varepsilon_0\omega$ |
|  | Susceptibility | $\chi$ | $i\sigma/\varepsilon_0\omega$ |
|  | Dielectric function | $\varepsilon$ | $1+\chi$ |
|  | Refractive index | $n$ | $\sqrt{\varepsilon}$ |
| Sheet response functions | Sheet conductivity | $\sigma^s$ | $\sigma d$ |
|  | Sheet susceptibility | $\chi^s$ | $\chi d$ |

In the following, we start with the well-known formulation of the optics of layered 3D material system (Section II B). We then consider the behavior in the limit of thin overlayers, defined as having a complex optical phase shift with a magnitude much less than unity. We show that the layer's optical response then appears only in the form of its sheet response function and present convenient expressions for the transmission and reflection coefficients of the system in this 2D limit (Section II C). If the sheet response of the layer is sufficiently weak, one can make use of a convenient linearized version of the 2D model (Section II D). Although these different results for the optical response of the system have all appeared previously in the literature, here we explicitly examine their relation to one another, the precise criteria for their validity, and their applicability to the inference of the material response of the thin layer from experimental measurements.

### II B. 3D MODEL

The problem of the optical response of a layer of thickness $d$ placed upon a semi-infinite substrate is well known. The complex reflection $r$ and transmission $t$ coefficients can be written (for arbitrary $d$) as [15]:

$$r = \frac{(1 - n_s)\cos\varphi - i(n_s - n^2)\varphi_0\left(\frac{\sin\varphi}{\varphi}\right)}{(1 + n_s)\cos\varphi - i(n_s + n^2)\varphi_0\left(\frac{\sin\varphi}{\varphi}\right)} e^{-2i\varphi_0} \tag{1}$$

$$t = \frac{2}{(1 + n_s)\cos\varphi - i(n_s + n^2)\varphi_0\left(\frac{\sin\varphi}{\varphi}\right)} e^{-i\varphi_0} \tag{2}$$

Here $n$ and $n_s$ are the (complex) refractive indices of the overlayer and the substrate, respectively. The two additional quantities appearing the relations are the optical phase shift $\varphi$ for the film of thickness $d$ and the corresponding vacuum phase shift $\varphi_0$: $\varphi = nk_0d = 2\pi nd/\lambda$ and $\varphi_0 = k_0d = 2\pi d/\lambda$, where $\lambda$ is the vacuum wavelength of light and $k_0$ is the corresponding vacuum wavevector. The (complex) phase shift $\varphi$ describes the change in the optical field in a single pass through the overlayer, while $\varphi_0$ is the corresponding phase evolution *in vacuo*. For the purpose of later discussions, we present some convenient expressions for $\varphi_0$ and $\varphi$ in Table 2.

We note that the relations (1) and (2) are written with respect to the coordinate system defined in Fig. 1, i.e., with the reflection and transmission coefficients describing the relation between the reflected and transmitted fields evaluated at the top of the substrate ($x = 0$) and compared to the value of the incident field evaluated in the same plane. For the incident and reflected fields, these values are extrapolated from the actual fields assuming that overlayer is absent. This definition is natural for describing the behavior of a thin film on a substrate and facilities comparison of the response for the case of a bare substrate. We note, however, that it differs from the usual convention for the reflection and transmission for an isolated thin film where the reflection coefficient is defined in terms of the reflected field measured from the left face of the slab compared to the incident field in the same plane, and the transmission coefficient relates the field at the right face of the slab to the field incident on the left face. While the substrate is assumed to be semi-infinite in our model, in an actual experiment with substrate of finite thickness, care needs to be taken to convert the measured quantities to the quantities in the model. Often, a wedged substrate or a confocal measurement scheme can be used to eliminate the reflection from the right surface of the substrate from the measured light intensity. In the transmission measurement, which is mostly performed beyond the right face of the substrate, light scattering at the right surface should be taken into account.

Experimentally, we measure the fraction of reflected or transmitted power of the light, *i.e.*, the corresponding reflectance $R$ or transmittance $T$ of the overlayer. These quantities are given directly in terms of the complex reflection and transmission coefficients in Eqs. 1-2 as

$$R = |r|^2 \qquad (3)$$

$$T = \text{Re}[n_s]|t|^2 \qquad (4)$$

$$A = 1 - R - T. \qquad (5)$$

In the above, we have also included the fraction of the incident optical power $A$ absorbed by the thin film, as dictated by energy conservation.

**Table 2.** Useful expressions for $\varphi_0$, $\varphi$ and $Z_0\sigma^s$. $Z_0 = 1/(\varepsilon_0 c) \approx 377\Omega$ is the impedance of free space.

|  | Expressed in $n, \varphi_0$ | Expressed in $\varepsilon, \varphi_0$ | Expressed in $n, d, \lambda$ |
|---|---|---|---|
| $\varphi_0$ | $\varphi_0$ | $\varphi_0$ | $2\pi(d/\lambda)$ |
| $\varphi$ | $n\varphi_0$ | $\sqrt{\varepsilon}\varphi_0$ | $2\pi n(d/\lambda)$ |
| $Z_0\sigma^s$ | $-i(n^2-1)\varphi_0$ | $-i(\varepsilon-1)\varphi_0$ | $-i2\pi(n^2-1)(d/\lambda)$ |

**II C. 2D MODEL**

The natural definition for the 2D limit for the optical response is one in which optical propagation effects in traversing the thin film are slight. We consequently consider the 2D limit of our problem to be defined by overlayer as satisfying the criterion of inducing a small optical phase shift, i.e., $|\varphi| \ll 1$. As we show explicitly below, in this limit the measurable optical response has no explicit dependence on the layer thickness, but can be expressed in terms of the optical sheet response of the layer.

We can obtain the results for optical response for the 2D model directly from the explicit expressions for the layered 3D system by taking the appropriate limit for a small optical phase shift in the film. In particular, in the relations for the reflection and transmission coefficients [Eqs. (1-2)] we linearize the trigonometric and exponential functions ($\cos \varphi \to 1$, $\sin \varphi / \varphi \to 1$, and $e^{-i\varphi_0} \to 1 - i\varphi_0$) in Eqs. 1-2. (In this paper we assume $\varphi_0$ to be small, as long as $|\varphi|$ is small, because $\varphi = n\varphi_0$ and $|n| \gtrsim 1$ for usual materials.) We then further simplify the relations in a way that is accurate to linear order in $\varphi_0$. Using the relations in Table 2, we can write the reflection and transmission coefficients in a form that depends only on the thin film's sheet response $\sigma^s = -i\chi^s \varepsilon_0 \omega$:

$$r = \frac{1 - n_s - Z_0 \sigma^s}{1 + n_s + Z_0 \sigma^s} \tag{6}$$

$$t = \frac{2}{1 + n_s + Z_0 \sigma^s} \tag{7}$$

Here $Z_0 = 1/(\varepsilon_0 c) \approx 377\Omega$ is the impedance of free space, and $Z_0 \sigma^s$ is a dimensionless quantity. In our simplification of the general result of the 3D model, we have assumed $\varphi_0$ and $\varphi$ to be small quantities, but allow $n^2 \varphi_0$ to be large and do not linearize with respect to this quantity. In this manner, we obtain a limit of the 3D response that does *not* require the normalized sheet conductivity of the layer $Z_0 \sigma^s = -i(n^2 \varphi_0 - \varphi_0)$ (Table 2) to be small. Consequently, the expressions can correctly predict strong overall optical response induced by the thin film, such as $T \ll 1$, as we illustrate below.

Although these relations can describe a strong optical response of the thin film, they are indeed still compatible with the lack of propagation effects, as one would expect in the limit of small optical phase shifts. This situation is signaled by the lack of any explicit dependence on the film thickness $d$. It is also evident from the form of the differential change induced by the presence of the thin film. Consider the change in the reflection $\Delta r = r - r_0$ and transmission $\Delta t = t - t_0$ coefficients with the corresponding coefficients for the bare substrate. From Eqs (6-7), we see that $\Delta r = \Delta t = \frac{-2Z_0\sigma^s}{(1+n_s)(1+n_s+Z_0\sigma^s)}$, i.e., the effect of the thin film on the optical fields is identical in the forward and reflected directions.

An alternative way to obtain the 2D model is by considering the electromagnetic boundary conditions of the thin layer. Under normal incidence, both the electric and magnetic fields are parallel to the surface; the boundary conditions are the continuity of electric field across the layer and a discontinuity of magnetic fields corresponding to the induced sheet current. [24] From these boundary conditions, we can directly obtain the transfer matrix for a 2D sheet, which may be used to analyze the effect of complex dielectric environment or the combination of multiple layers. The transfer matrices relate the electric and magnetic fields at the two boundaries (I and II) of the medium: $\begin{pmatrix} E_I \\ H_I \end{pmatrix} = M^s \begin{pmatrix} E_{II} \\ H_{II} \end{pmatrix}$. [15]

$$M^s = \begin{pmatrix} 1 & 0 \\ \sigma^s & 1 \end{pmatrix} \quad (8)$$

The transfer matrix can then be combined with the refractive index of the substrate to obtain Eqs. 6-7. The 2D expressions for the reflectance, transmittance, and absorption, obtained by inserting Eqs. 6-7 into 3-5, are listed below.

$$R = \left| \frac{1 - n_s - Z_0\sigma^s}{1 + n_s + Z_0\sigma^s} \right|^2 \quad (9)$$

$$T = \text{Re}[n_s] \left| \frac{2}{1 + n_s + Z_0 \sigma^s} \right|^2 \qquad (10)$$

$$A = \text{Re}[Z_0 \sigma^s] \left| \frac{2}{1 + n_s + Z_0 \sigma^s} \right|^2 \qquad (11)$$

**II D. LINEARIZED 2D MODEL**

Within the 2D limit, an important regime is when the presence of the thin sheet can be treated as a perturbation. In this regime, the optical quantities such as $r$ and $t$ vary linearly with the sheet response of the thin layer to good approximation. Their linearized expressions are obtained by expanding the corresponding expressions in the full 2D model to linear order in $\sigma^s$:

$$r = r_0 \left( 1 - \frac{2}{1 - n_s^2} Z_0 \sigma^s \right) \qquad (12)$$

$$t = t_0 \left( 1 - \frac{1}{1 + n_s} Z_0 \sigma^s \right) \qquad (13)$$

Mathematically, the linearization requires $Z_0 \sigma^s$, and hence $n^2 \varphi_0$ (Table 2) to be small. This additional requirement distinguishes the linearized 2D model from the more general form. While the exact expression for $Z_0 \sigma^s$ in terms of $\varphi_0$ is $(n^2 - 1)\varphi_0$, here and in the rest of this paper we use $n^2$ and $n^2 - 1$ interchangeably for simplicity. For an example, we may use the approximation $Z_0 \sigma^s \approx n^2 \varphi_0$. We note that the linearized 2D model is also the rigorous limit of the first-order expansion of the 3D model in $d$.

Using Eqs. (12-13), we obtain the reflection contrast $\Delta R/R \equiv (R - R_0)/R_0$, transmission contrast $\Delta T/T \equiv (T - T_0)/T_0$, and absorbance $A$ (Eqs. 14-16). $R_0$ and $T_0$ denote the reflectance and transmittance from the bare substrate. These optical contrasts can often be conveniently measured in the far-field and they are also referred to as the differential reflectance and differential transmittance.

[1,2,16] Various forms of the linearized 2D model appeared in earlier literatures on the optics of thin films. [16,17]

$$\Delta R/R = \text{Re}\left[\frac{4}{n_s^2 - 1} Z_0 \sigma^s\right] \tag{14}$$

$$\Delta T/T = -\text{Re}\left[\frac{2}{1 + n_s} Z_0 \sigma^s\right] \tag{15}$$

$$A = \frac{4}{|1 + n_s|^2} \text{Re}[Z_0 \sigma^s] \tag{16}$$

As can be seen in Eqs. 14-15, for non-absorbing substrate, *i.e.* when $n_s$ is real, the optical contrasts are directly proportional to $\text{Re}[Z_0 \sigma^s]$ in the linear approximation.[17] Here the total reflected and transmitted field can be viewed to linear order as the coherent sum of the field corresponding to bare substrate and the radiation from the 2D sheet. When the substrate is non-dissipative, the radiation from the 2D sheet due to $\text{Re}[\sigma^s]$ is in phase (or 180° out of phase) with that from the substrate. On the other hand, radiation due to $\text{Im}[\sigma^s]$ of the 2D sheet belong to the quadrature component. As a result, only $\text{Re}[\sigma^s]$ (or $\text{Im}[\varepsilon]$) is present in the linear optical contrast. When the substrate is allowed to be dissipative, however, the reflection and transmission contrasts are in general sensitive to both $\text{Re}[\sigma^s]$ and $\text{Im}[\sigma^s]$ to linear order. We further note from Eq. 16 that the absorbance is strictly proportional to the real part of $\sigma^s$ in the linear approximation, regardless of the realness of the substrate refractive index. This can be intuitively understood as the following: the absorbance (or energy dissipation) is proportional to the product of the absolute square of the local field and the real part of the sheet conductivity. Even though the local field can be affected to linear order in the conductivity, the correction in absorbance is in the second and higher orders.

### III. MEASUREMENT OF OPTICAL RESPONSE FUNCTIONS
#### III A. GENERAL CASE

By measuring the reflection or the transmission coefficient, one can directly invert Eq. 6 or Eq. 7 to obtain the complex sheet conductivity. If only the intensity of light is measured, one can in principle obtain both the reflectance and transmittance, and then solve for the complex sheet conductivity using Eqs. 8-9. This method is not always applicable, however. For an example, in a common case of a conductive layer on a transparent substrate, both the reflectance and transmittance are primarily sensitive to only the real part of sheet conductivity (Eqs. 14-15). As a result, experimental uncertainties can lead to significant error in the determined imaginary part of sheet conductivity. Hence an alternative method is to make use of the Kramers-Kronig relations between the real and imaginary parts. Here a trial complex material response function is allowed to vary under the constraint of Kramers-Kronig relation until the modeled optical response matches with the measurement. [3-5] While complete information of the complex material response function can be extracted, the Kramers-Kronig analysis requires knowledge of the material optical response outside of the measurement window.

### III B. LINEAR APPROXIMATION

To first order, the reflectance and transmittance contrasts are only sensitive to $\text{Re}[\sigma^s]$ for a thin layer on a transparent substrate (Eqs. 14-15). Hence in the linear approximation, $\text{Re}[\sigma^s]$ can be directly inferred from a measurement of the optical contrast. This method has been applied to graphene in both reflection [1] and transmission [2] measurement configurations. To obtain the complete complex response function, additional measurements can be performed. For example, one can measure the optical contrast on a second substrate, where the optical contrast dependents strongly on the imaginary part of the sheet conductivity. The two measurements thus provide two non-degenerate sets of equations at each frequency, from which the complex response function is readily solvable. While the Kramers Kronig analysis can also be used in the linear model, the advantage of the simplified computation by the linear model is diminished and one would then prefer the more accurate model (Section IIIA).

## IV. APPLICABILITY OF THE 2D MODELS

### IV A. 2D MODEL

The criterion for the thin layer approximation is $|\varphi| \ll 1$. Single atomic layers have thicknesses about 3 orders of magnitudes smaller than the optical wavelength, giving rise to $\varphi_0 = 2\pi d/\lambda$ of less than 0.01. On the other hand, $|\varepsilon|$ is typically less than 100 in the visible region (Fig. 2). Correspondingly, $|\varphi| = |\sqrt{\varepsilon}\varphi_0| < 0.1$ (Fig. 2c and f). Hence we expect the 2D model to work well for atomic layers. To verify this expectation, we consider monolayer $WS_2$ as a representative atomically thin layer. Monolayer $WS_2$ has strong optical response in the visible region, particularly near the excitonic resonance at around 2.0 eV, as shown in Figs. 2a-b. Figs. 3a-c compare the 2D and 3D model predictions of the optical absorbance, reflectance contrast, and transmittance contrast spectra of monolayer $WS_2$ on fused silica substrates. Excellent agreement of the 2D model with the 3D model in the visible region is evident from the accurate alignment of the spectra produced from the two models. Since $WS_2$ is a semiconductor, the magnitude of its dielectric function below the band gap in the infrared is of similar magnitudes to that in the visible. We note that the 2D model will breakdown for even single atomic layers as long as $|\varepsilon|$ is large enough such that $|\varphi| = |\sqrt{\varepsilon}\varphi_0|$ is comparable or larger than 1. These scenarios can potentially be achieved with sharp excitonic resonances.

To check the validity of 2D approximation for conductive atomic layers, we consider doped graphene as a representative example. We model the optical conductivity of doped graphene as a sum of intraband (free-carrier) and interband responses [25]. Details of the conductivity model are provided in the Supplemental Material, Section III. The magnitude of graphene's dielectric function increases rapidly in the infrared due to the free-carrier response. The increase in dielectric function, however, is counter-balanced by the increase in wavelength, so that $|\varphi| = \left|2\pi\sqrt{\varepsilon}d/\lambda\right|$ remains small throughout the infrared range (Fig. 2f). We illustrate the accuracy of the 2D model for graphene from DC to infrared, and up to UV using the optical spectra in Figs. 3d-f.

The 2D approximation is robust even beyond single atomic layers. The values of $|\varphi|$ for graphene and $WS_2$ in Figs. 2c and f suggests that the 2D model should hold up to 5-10 atomic layers. (Assuming that the dielectric function does not vary with layer thickness, $\varphi$ is then proportional to the number of layers.) This is validated for multi-layers of $WS_2$ in the Supplemental material, Section IV. Apart from van der Waals layers, we also consider thin gold layers as a representative case for metallic thin layers. Good conductors have strong DC response. For gold, $-\text{Re}[\varepsilon]$ reaches over $10^5$ and $\text{Im}[\varepsilon]$ diverges, as shown in Figs. 2g-h. Here the dielectric function of gold is modeled by a Drude response fitted to experimental data.[26] (Further discussions of the behavior of Drude response is given in the Supplemental Material, Section II.) The transmission contrast given by a 10 nm gold layer approaches 100% ($T$ approaches 0) near DC (Fig. 3i). With such strong response, Figs 3g-i show that the 2D model remains accurate, which is consistent with the value of $|\varphi|$ in Fig 2i.

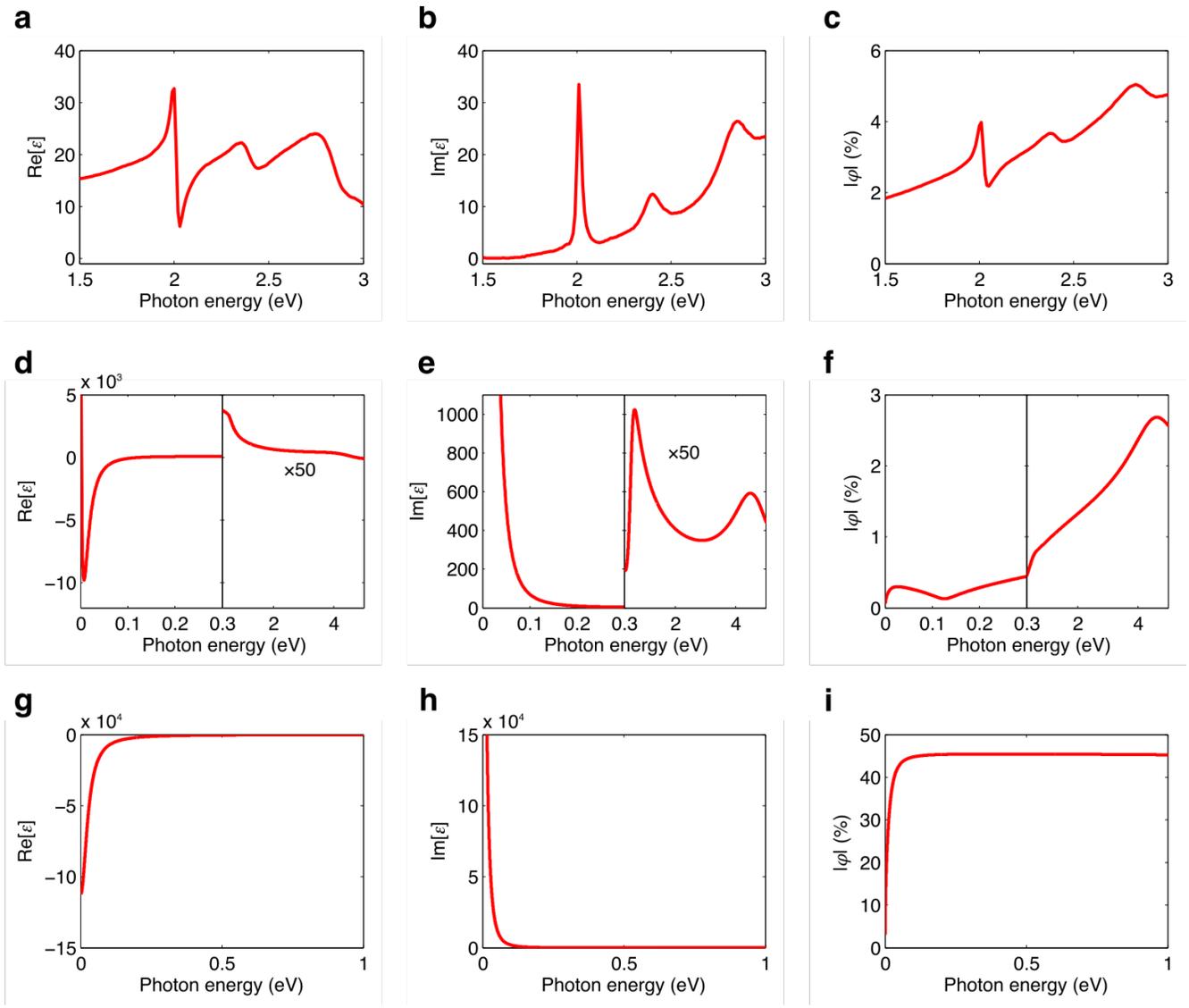

**Figure 2**. Real and imaginary parts of the dielectric function $\varepsilon$ and $|\varphi|$ for monolayer WS$_2$ **(a-c)**, monolayer graphene **(d-f)** and 10 nm of gold **(g-i)**.

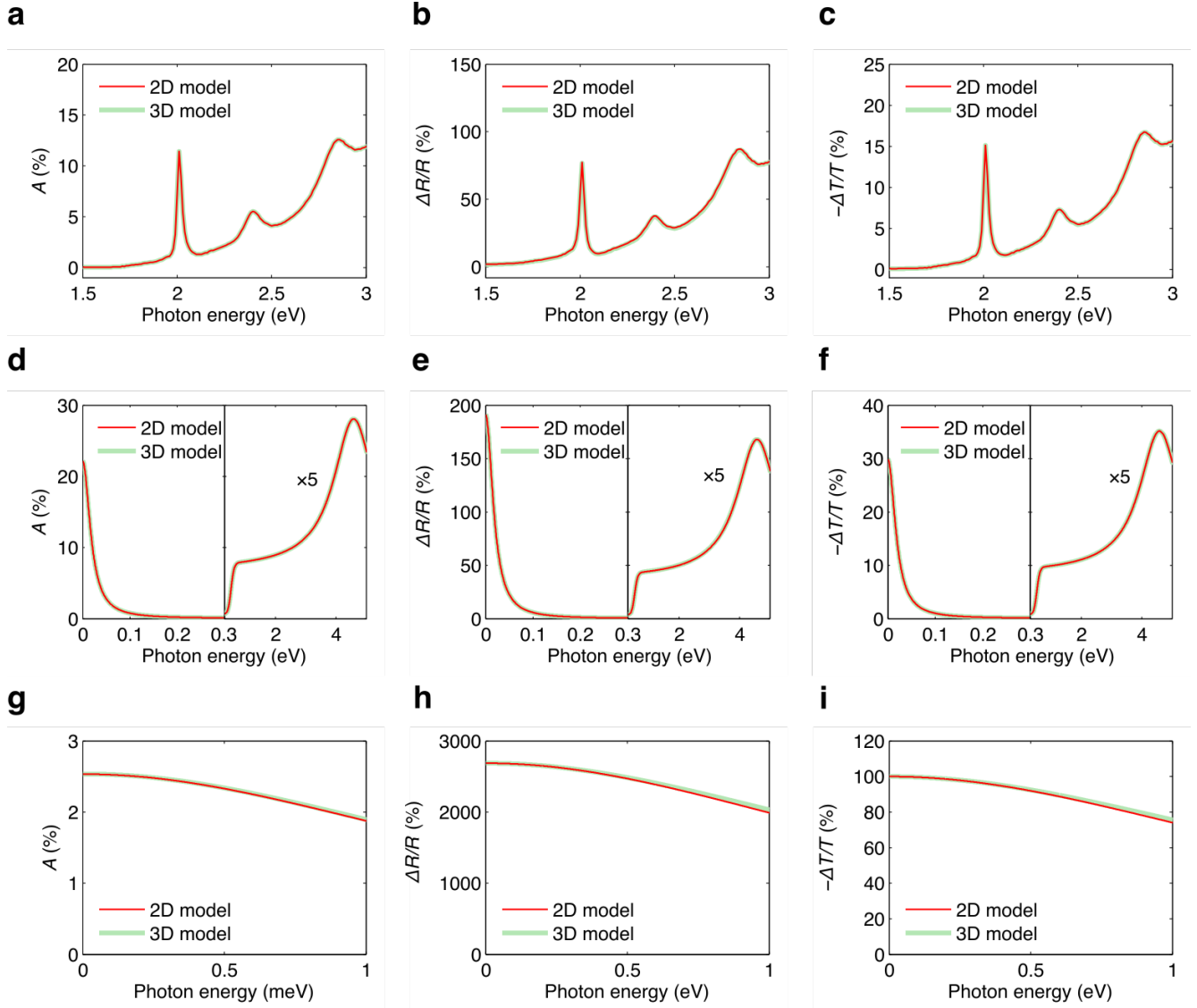

**Figure 3**. Absorbance, reflectance contrast and transmittance contrast for monolayer WS$_2$ **(a-c)**, monolayer graphene **(d-f)**, and 10 nm thick gold layer **(g-i)**, all on fused silica substrate.

## IV B. LINEARIZED 2D MODEL

According to the derivation of the linearized 2D model, linear approximations to $r$, $t$, $A$, and $\Delta T/T$ (Eqs. 12-13 and 15-16) require that $|Z_0 \sigma^s| \ll |1 + n_s|$, or, equivalently, $|\varepsilon| \ll |1 + n_s|\varphi_0^{-1}$. For $\Delta R/R$ (Eq. 14), the condition of $|Z_0 \sigma^s| \ll |1 - n_s|$ (or $|\varepsilon| \ll |1 - n_s|\varphi_0^{-1}$) needs to be satisfied additionally. For usual substrate with $n_s$ sufficiently different from 1, we can ignore the pre-factors and

write a simplified criterion for all the physical quantities considered in the linearized 2D model: $|Z_0\sigma^s| \ll 1$ (Table 3). As shown in Figs. 4a and d, $|Z_0\sigma^s| < 0.5$ for the representative atomic layers on fused silica in the spectral ranges of concern. The real and imaginary parts of $Z_0\sigma^s$ are shown in Figs. 4b-c and 4e-f for reference. We then expect that the linear approximation to apply reasonably well. We plot in Figs. 5a-f the absorbance, reflectance contrast and transmittance contrast of monolayer $WS_2$ and graphene calculated by the full 2D model and the linearized version. Indeed, the linear approximation is good for these atomically thin layers in most parts of the spectral range. For $WS_2$, the deviation is most significant near the band gap of 2.0 eV and in the blue range due to interband transitions (Fig. 5a-c). For graphene, the deviation is most significant near DC, where the Drude response peaks (Fig 5d-f). In the case of a 10 nm gold layer, $|Z_0\sigma^s| > 4$ in the whole infrared region (Fig. 4g), and consequently the linear approximation completely breaks down – the optical properties of the gold layer predicted by the linearized 2D model disagree with those predicted by the exact 2D model (Figs. 5g-i). In particular, the linearized 2D model absurdly predicts an absorption that is larger than 100% below about 0.3 eV. We should emphasize that while the linearized 2D model completely breaks down in this example, the full 2D model remained accurate (Figs. 3g-i).

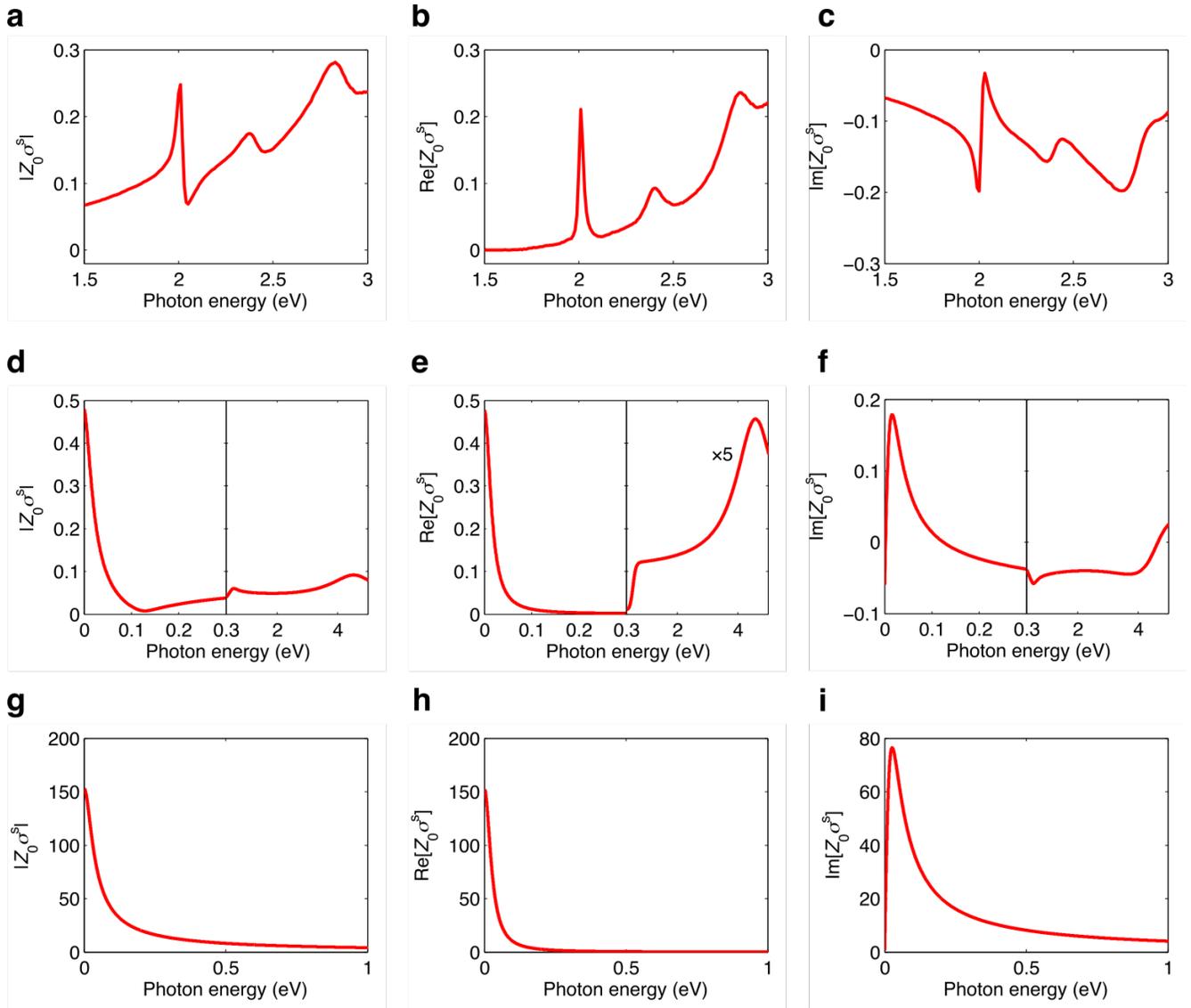

**Figure 4.** Absolute value and the real and imaginary parts of $Z_0\sigma^s$ for monolayer WS$_2$, monolayer graphene and 10 nm thick gold layer.

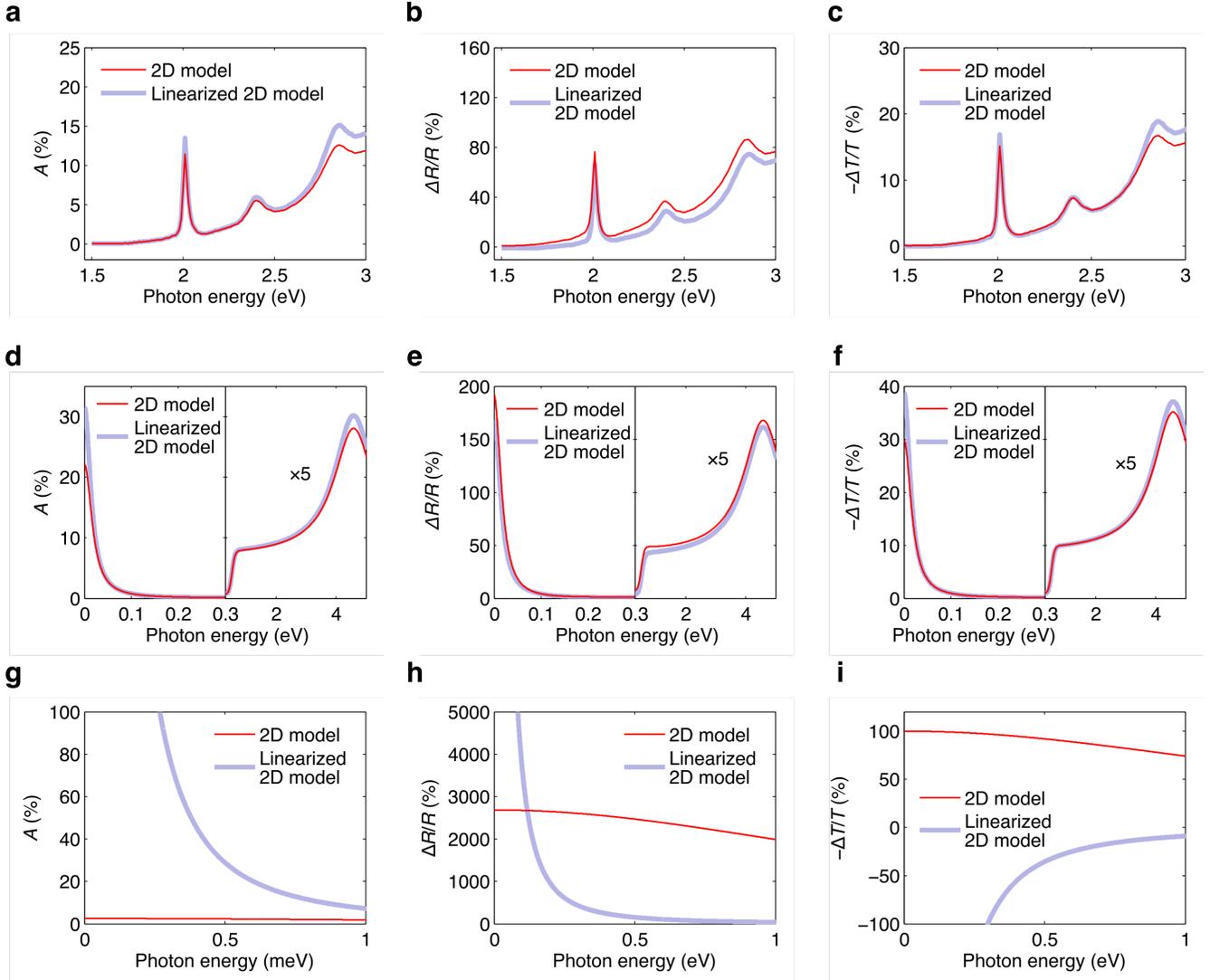

**Figure 5**. Absorbance, reflectance contrast and transmittance contrast of monolayer WS$_2$ **(a-c)**, monolayer graphene **(d-f)** and 10 nm gold layer **(g-h)** obtained from the 2D model and from the linearized 2D model.

## IV C. SUMMARY OF APPLICABILITY

Here we summarize the applicability of the 2D, linearized 2D and 3D models, and express them in terms of the optical thickness $\varphi$. In Section IV B, we have shown that the linearized 2D model is applicable when the presence of the layer only slightly modifies the optical response of the substrate, and the corresponding analytical expression is $|Z_0 \sigma^s| \ll 1$. This is equivalent to $|\varphi| \ll 1/|n|$ according to Table II. In Section IV A, we showed that the 2D model is applicable when the optical thickness of the layer is small, *i.e.*, $|\varphi| \ll 1$. On the other hand, the 3D model is generally applicable for all values of $|\varphi|$. To further illustrate the relations of the applicability of the 3 models, we indicate their ranges of applicability on the axis of increasing $|\varphi|$ (Fig. 6). When $|\varphi| \ll 1/|n|$, all 3 models are applicable, while the linear model provides the simplest mathematical expressions. In the region where $|\varphi| \gtrsim 1/|n|$, but $|\varphi| \ll 1$, the thin layer can no longer be treated as a perturbation and the linearized model is no longer valid. While the layer is still optically thin, the 2D model provides the most concise description of the thin layer, but the 3D model is also applicable. When $|\varphi| \gtrsim 1$, the 3D model becomes the only appropriate model for the system. Note that even for a layer that is physically thin, *i.e.*, $\varphi_0 \ll 1$, it is possible that $|\varphi| \gtrsim 1$ and $|\varphi| \gtrsim 1/|n|$ when $n$ is sufficiently large. Here we have expressed the ranges of applicability using the notation of 'much smaller'. Physically, applicability of a model should be determined by the allowed error generated in the model. Hence the meaning of 'much smaller' should be understood in the sense that the quantity on the left hand side is so much smaller than the right hand side such that the error introduced in the model is acceptable.

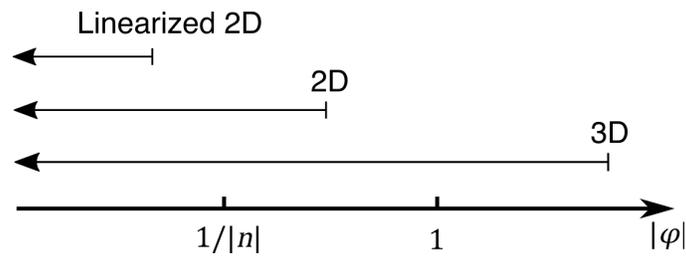

**Figure 6.** Summary of the applicability of the models in terms of the magnitude of the optical thickness, $|\varphi|$. Here we assume the normal response of a material where $|n| > 1$. The schematic illustrates the increasing range of applicability from the linearized 2D model to the 2D model, and to the 3D model. The values of $|\varphi|$ that provide the bounds of applicability are labeled on the horizontal axis. The bottom axis indicates increasing magnitude of $\varphi$, but does not correspond to a quantitative scale.

## IVD. EXPERIMENTAL CRITERIA OF APPLICABILITY

The applicability criteria of the 2D model and its linearized version have been obtained considering the approximations used in formulating these models, and known dielectric response is assumed. In practice, however, the optical properties of the material under test may not be known *a priori*. Hence we discuss below a set of experimental criteria that can be verified from measured optical fields or intensities.

We have seen earlier that the change in transmission and reflection induced by a thin layer is equal in the 2D limit, *i.e.*, $\Delta t = \Delta r$. In fact, when the physical thickness of the layer is small compared to the reduced wavelength of light, *i.e.*, $d \ll \lambda/2\pi$, $|(\Delta t - \Delta r)/t| \ll 1$ (Table 3) is necessary for the applicability of the 2D model, and if it is true over a broad range of frequencies, it becomes also sufficient (see Supplemental Material, Section VII). For the linearized 2D model, we highlight $\Delta R/R \ll 1$ and $\Delta T/T \ll 1$ as a set of experimental criteria for the linear approximation of $\Delta R/R$ and $\Delta T/T$, respectively. For non-absorbing substrate and superstrate, $\Delta T/T \ll 1$ is necessary and sufficient for the validity of the linearized 2D model; $\Delta R/R \ll 1$ is necessary and sufficient, except for certain anomalous cases (see Supplemental Material, Section VIII). The rationale behind the experimental criteria for the linearized model is that small optical contrast indicates weak material response.

## V. ACCURACY OF THE 2D MODELS

Accuracies of the 2D models are determined by the leading terms dropped by the approximations from the 3D model. For the 2D model, propagation of the errors in the approximations of $\cos\varphi \to 1$, $\sin\varphi/\varphi \to 1$, and $e^{-i\varphi_0} \to 1 - i\varphi_0$ leads to accuracies proportional to $|\varphi|^2$. Accuracies for the linearized 2D model are obtained in a similar fashion and are proportional to

$|Z_0\sigma^s|^2$, or $\varepsilon|\varphi|^2$. Table 3 lists the accuracies of the full 2D model and the linearized 2D model without pre-factors. Complete expressions for the fractional accuracies with pre-factors are given in the SI.

**Table 3.** Applicability and accuracy of the 2D model and linearized 2D model. Pre-factors are omitted in the expressions below for simplicity. Full expressions are provided in the Supplemental Material, Section IX.

|  |  | 2D model |  | Linearized 2D model |
|---|---|---|---|---|
| Applicability | Bulk criteria: | $|\varphi| \ll 1$ or $|n| \ll \varphi_0^{-1}$ | Bulk criteria: | $|\varphi| \ll |n|^{-1}$ or $|n| \ll \varphi_0^{-1/2}$ |
|  | Sheet criterion: | $|Z_0\sigma^s| \ll |n|$ | Sheet criterion: | $|Z_0\sigma^s| \ll 1$ |
|  | Experimental criterion: | $|(\Delta t - \Delta r)/t| \ll 1$ | Experimental criteria: | $\Delta T/T \ll 1$ or $\Delta R/R \ll 1$ |
| Accuracy | For $X = r, R, t, T$: | $|\varphi|^2 X$ | For $X = r, R, t, T$: | $|n\varphi|^2 X$ or $|Z_0\sigma^s|^2 X$ |
|  | For $A$, $\Delta T/T$, and $\Delta R/R$: | $|\varphi|^2$ | For $A$, $\Delta T/T$, and $\Delta R/R$: | $|n\varphi|^2$ or $|Z_0\sigma^s|^2$ |

## VI. IMPLICATIONS OF THE 2D MODEL

The 2D model can be compared to a naïve model where scattering at the interfaces are ignored. For the ease of analysis, we consider the case of a thin layer in free-space. In the naïve model, the change in transmission coefficient $\Delta t = e^{i\varphi} - e^{i\varphi_0} \approx i(n-1)\varphi_0$, where in the last step the 2D approximation of $|\varphi| \ll 1$ is used. In the 2D model that correctly takes into account the scattering at interfaces, the change in transmission coefficient is $\Delta t = \frac{n+1}{2+Z_0\sigma^s} i(n-1)\varphi_0$. Ignorance of scattering at the boundaries gives rise to a factor of $\frac{n+1}{2+Z_0\sigma^s}$ difference in the estimated change in the transmission coefficient. In addition, we compare the intensity contrast in the two models. In the naïve model, the transmission contrast (attenuation) is directly proportional to $\text{Im}[n]$: $\Delta T/T = e^{-2\text{Im}[n]\varphi_0} - 1 \approx -2\text{Im}[n]\varphi_0$. This expression is the same as the Beer's law. On the other hand, the 2D model gives $\frac{\Delta T}{T} = \frac{-4\text{Re}[Z_0\sigma^s] - |Z_0\sigma^s|^2}{|2+Z_0\sigma^s|^2}$ and in the linear approximation, $\frac{\Delta T}{T} = -\text{Re}[Z_0\sigma^s] = -\text{Im}[\varepsilon]\varphi_0$. Since $\text{Im}[\varepsilon] = 2\text{Re}[n]\text{Im}[n]$ by definition, naïve (and incorrect) application of the Beer's law would yield a result that differs from the actual attenuation by a factor of $\text{Re}[n]$ compared to the linearized 2D model. The discrepancy is again attributed to the dielectric discontinuities presenting at the boundaries of the thin layer. These discontinuities introduce multiple reflections within the layer and modify the transmission contrast when $\text{Re}[n]$ differs from 1. The above analysis also shows that large dielectric discontinuities can substantially reduce the required thickness of a material to obtain a certain amount of transmission attenuation.

In addition to the implications for transmission attenuation, an interesting observation from the 2D model is that a 2D sheet can be strongly absorptive and optically thin at the same time. This would allow the realization of ultrathin optical absorbers.[18,27,28] In our graphene example, the absorption is more than 20% near DC while the 2D model still captures accurately its absorption (Fig. 3g). By taking the derivative of the absorption formula (Eq. 10) with respect to $\sigma^s$ and assuming that the sheet conductivity to be purely real, we find the maximum absorption to be $A_{max} = 1/(n_s + 1)$, occurring at $\sigma^s = (n_s + 1)/Z_0$. Hence for a free-standing layer ($n_s = 1$), the maximum absorption is 50% and the required sheet conductivity is $2/Z_0$. We further note that according to the expression for $A_{max}$, 100% absorption should be attainable with $n_s = 0$. While $n_s = 0$ is uncommon in bulk materials, it can be shown that a good mirror at a distance of $\lambda/4$ from the 2D layer is equivalent to an effective substrate with refractive index of $n_s = 0$ immediately next to the 2D layer for normally incident light. The equivalency will be proved in next section.

## VII. GENERALIZATION OF THE OPTICAL MODELS

We make three generalizations to the 2D model to include anisotropic 2D layer, heterostructures, and substrate with multiple layers. For a thin layer with in-plane anisotropy, the incident electric field can be projected onto the two in-plane principal axes. The analysis can then be carried out for each component as the isotropic case. The resultant reflected and transmitted fields will then be the vector sum of the two principal components. Due to the anisotropy of the material optical response, the reflected and transmitted fields in general have different polarizations as the excitation field.

For heterostructures, one can apply the 2D transfer matrix formalism to obtain the total response from the individual components [15]. As can be seen by multiplying the 2D transfer matrices, the total response of a heterostructure is simply the summation of the responses from each constituent layer. (Here we assume that the individual components are isotropic.) More explicitly, for a heterostructures that contains $m$ layers each having an in-plane sheet conductivity $\sigma_i^s$ ($i$ = 1, 2, 3, …, $m$), the effective optical conductivity of the heterostructures is $\sigma^s = \sigma_1^s + \sigma_2^s + \sigma_3^s +, ..., +\sigma_m^s$. If the heterostructure contains anisotropic layers, however, a full analysis considering the polarization is needed.

Finally, for a general substrate, such as oxidized silicon, one can combine the 2D transfer matrix (Eq. 11) of the thin layer with the 3D transfer matrix of the substrate by matrix multiplication to analyze the optical properties of composite systems.[15] For the analysis of reflection and absorption, one can simply find an effective homogeneous substrate with refractive index $n_s$ that produces the same reflection coefficient $r_s$ (bare substrate) as the general substrate under consideration. The relation between the effective $n_s$ and $r_s$ is $r_s = (1 - n_s)/(1 + n_s)$. This equivalence can be understood considering an infinitesimally thin vacuum layer between the thin material and the substrate. The infinitesimally thin vacuum layer does not change the optics of the system. Since the effect of the substrate comes only into the problem through its reflection coefficient with vacuum above, optics of the thin layer (except for transmission) will be the same for the effective substrate and the actual substrate. The expressions for reflection and absorption (Eqs. 1, 6, 8, 10, 12, 14, 16) can thus be expressed with the substitution $n_s \to (1 - r_s)/(1 + r_s)$, which is derived from the expression of reflection coefficient for a homogeneous substrate above.

Knowing how to use an effective refractive index to model the substrate, we can now show that $n_s = 0$ can be achieved effectively by placing a perfect mirror at a distance of $\lambda/4$ from the origin of reference, for the realization of a total absorber discussed in the previous section. A perfect mirror has $r_s = -1$ when the origin of reference is right on top of the mirror. When the origin of reference is placed $\lambda/4$ away from the mirror, the effective reflection coefficient would be $r_s = (-1)e^{i\pi} = 1$. We plug $r_s = 1$ into $(1 - r_s)/(1 + r_s)$ and obtain the effective substrate refractive index $n_s = 0$. Thus we proved that a perfect mirror placed $\lambda/4$ away from the mirror provides an effective substrate with $n_s = 0$ at wavelength $\lambda$. This is the underlying principle for the Salibury screen, a way of reducing the reflection of radio waves from a surface. [19]

This paper is focused on plane waves and normal incidence geometry. An extension of the current framework to oblique incidence is of general interest, and will be particularly useful for ellipsometry measurements.[6-14] This extension, however, is beyond the scope of this paper.

## VIII. CONCLUSION

In conclusion, we have described models for the optics of thin layers. Accuracy and applicability of the 2D model and its linearized version are analyzed by comparing to the 3D model. The 2D model is shown to be accurate over a broad frequency range from DC to near UV for typical atomically thin layers. Identical change in the reflection coefficient and transmission coefficients is given as an experimental criterion for the applicability of the 2D model. Small absorbance and optical contrasts are demonstrated as a set of practical criteria for the applicability of the linearized 2D model. We discussed how to measure the optical response functions of a thin layer using the developed model. The 2D model implies significantly less material than required by the Beer's law in achieving certain amount of transmission attenuation. The model also yields the conditions for realizing half and total absorption. At the end, the optical models are generalized for in-plane anisotropy, heterostructures, and general substrates.

## ACKNOWLEDGEMENT

The authors would like to acknowledge the helpful discussions with the members of Heinz group.